\documentclass{article}

\usepackage{microtype}
\usepackage{graphicx}
\usepackage{subfigure}
\usepackage{booktabs} 
\usepackage[skip=0pt]{caption}

\usepackage{xcolor}

\usepackage{hyperref}
\usepackage{CJKutf8}

\usepackage[accepted]{icml2023}

\usepackage{amsmath}
\usepackage{amssymb}
\usepackage{mathtools}
\usepackage{amsthm}

\usepackage[capitalize,noabbrev]{cleveref}

\theoremstyle{plain}

\theoremstyle{definition}

\theoremstyle{remark}

\usepackage[textsize=tiny]{todonotes}

\icmltitlerunning{Generating XP spectra with unsupervised learning}

\begin{document}

\twocolumn[
\icmltitle{Closing the stellar labels gap: \\
           An unsupervised, generative model for \textit{Gaia} BP/RP spectra}



\icmlsetsymbol{equal}{*}

\begin{icmlauthorlist}
\icmlauthor{Alexander Laroche}{toronto_astro}
\icmlauthor{Joshua S. Speagle (\begin{CJK*}{UTF8}{gbsn}沈佳士\end{CJK*})}{toronto_astro,toronto_stats,dunlap,dsi}
\end{icmlauthorlist}

\icmlaffiliation{toronto_astro}{David A. Dunlap Department of Astronomy \& Astrophysics, University of Toronto, 50 St George St, Toronto, ON M5S 3H4, Canada}
\icmlaffiliation{toronto_stats}{Department of Statistical Sciences, University of Toronto, 9th Floor, Ontario Power Building, 700 University Ave, Toronto, ON M5G 1Z5, Canada}
\icmlaffiliation{dunlap}{Dunlap Institute for Astronomy \& Astrophysics, University of Toronto, 50 St George Street, Toronto, ON M5S 3H4, Canada}
\icmlaffiliation{dsi}{Data Sciences Institute, University of Toronto, 17th Floor, Ontario Power Building, 700 University Ave, Toronto, ON M5G 1Z5, Canada}

\icmlcorrespondingauthor{Alexander Laroche}{alex.laroche@mail.utoronto.ca}

\icmlkeywords{Fundamental parameters of stars, Spectroscopy (1558), Astronomy data analysis, Astrostatistics, Neural networks}

\vskip 0.3in
]



\printAffiliationsAndNotice{}  

\begin{abstract}
The recent release of 220+ million BP/RP spectra in \textit{Gaia} DR3 presents an opportunity to apply deep learning models to an unprecedented number of stellar spectra, at extremely low-resolution. The BP/RP dataset is so massive that no previous spectroscopic survey can provide enough stellar labels to cover the BP/RP parameter space. We present an unsupervised, deep, generative model for BP/RP spectra: a \textit{scatter} variational auto-encoder. We design a non-traditional variational auto-encoder which is capable of modeling both $(i)$ BP/RP coefficients and $(ii)$ intrinsic scatter. Our model learns a latent space from which to generate BP/RP spectra (scatter) directly from the data itself without requiring any stellar labels. We demonstrate that our model accurately reproduces BP/RP spectra in regions of parameter space where supervised learning fails or cannot be implemented.
\end{abstract}


\section{Introduction}\label{sec:intro}

Data-driven models applied to high-resolution spectroscopy can produce very precise measurements of stellar labels: effective temperature $T_{\rm eff},$ surface gravity $\log g$ and metallicity $\rm [Fe/H]$. Seminal work done by \citet{Cannon-15} showed that a data-driven generative model with 2nd-order polynomials in stellar labels, called \textit{The Cannon}, can produce stellar label estimates which are as accurate as the physics-driven APOGEE pipeline \citep[ASPCAP:][]{APOGEE-22}. Recent work done by the LAMOST Survey \citep[$R\sim10^3$;][]{LAMOST-22} has demonstrated that medium-resolution spectroscopy can provide stellar label estimates which rival their high-resolution counterparts \citep{Wang++22}.

Data-driven approaches are particularly crucial for the 220+ million flux-calibrated, low-resolution spectra ($R\sim100$) in the recent \textit{Gaia} Data Release 3 \citep[GDR3,][]{GDR3-22}. These spectra are a combination of measurements from the Blue Photometer (BP) and Red Photometer (RP) \textit{Gaia} instruments which span 330-1050 nm. Spectroscopic stellar label estimates have thus far largely been limited to high-resolution ground-based surveys. For instance, APOGEE is heavily biased towards giant stars, disproportionately targets stars in the Galactic disk, and only contains $\sim10^5$ spectroscopic observations. The \textit{Gaia} XP spectra do not suffer from these shortcomings; \textit{Gaia} has excellent spatial coverage, targets stellar populations beyond giants, and contains more stars than APOGEE by 3 orders of magnitude. The evident shortcoming of the Gaia XP spectra is the low spectral resolution.

XP spectra analyses have only recently begun to be undertaken; uncovering carbon-enhanced metal-poor (CEMP) stars \citep{Lucey++22} and revealing the metal-poor Galactic center \citep{POH-22}. To date, the only generative model for XP spectra is that of \citet{Zhang++23}; a supervised deep learning model. Although \citet{Zhang++23} did present a stellar labels catalog for all 220 million stars in the XP dataset, a large fraction of them are unreliable. Specifically, their LAMOST training catalog (which only cross-matches $\sim1\%$ of the XP spectra) does not contain enough stellar labels for white/M dwarfs, B stars, high-extinction supergiants, etc. and as such their supervised model performs poorly when applied to these stellar populations. To date, no unsupervised generative models exist. An unsupervised model has the potential to fill the ‘stellar labels gap’ of \citet{Zhang++23}. By combining supervised and unsupervised approaches, it will be possible to develop a data-driven, generative model which accurately predicts the entire XP dataset. To that end, we present the first unsupervised, generative model for XP spectra; a \emph{scatter} variational auto-encoder. 

\section{Methods}\label{sec:meth}

\begin{figure*}[hbt!]
    \centering
    \includegraphics[width=\textwidth]{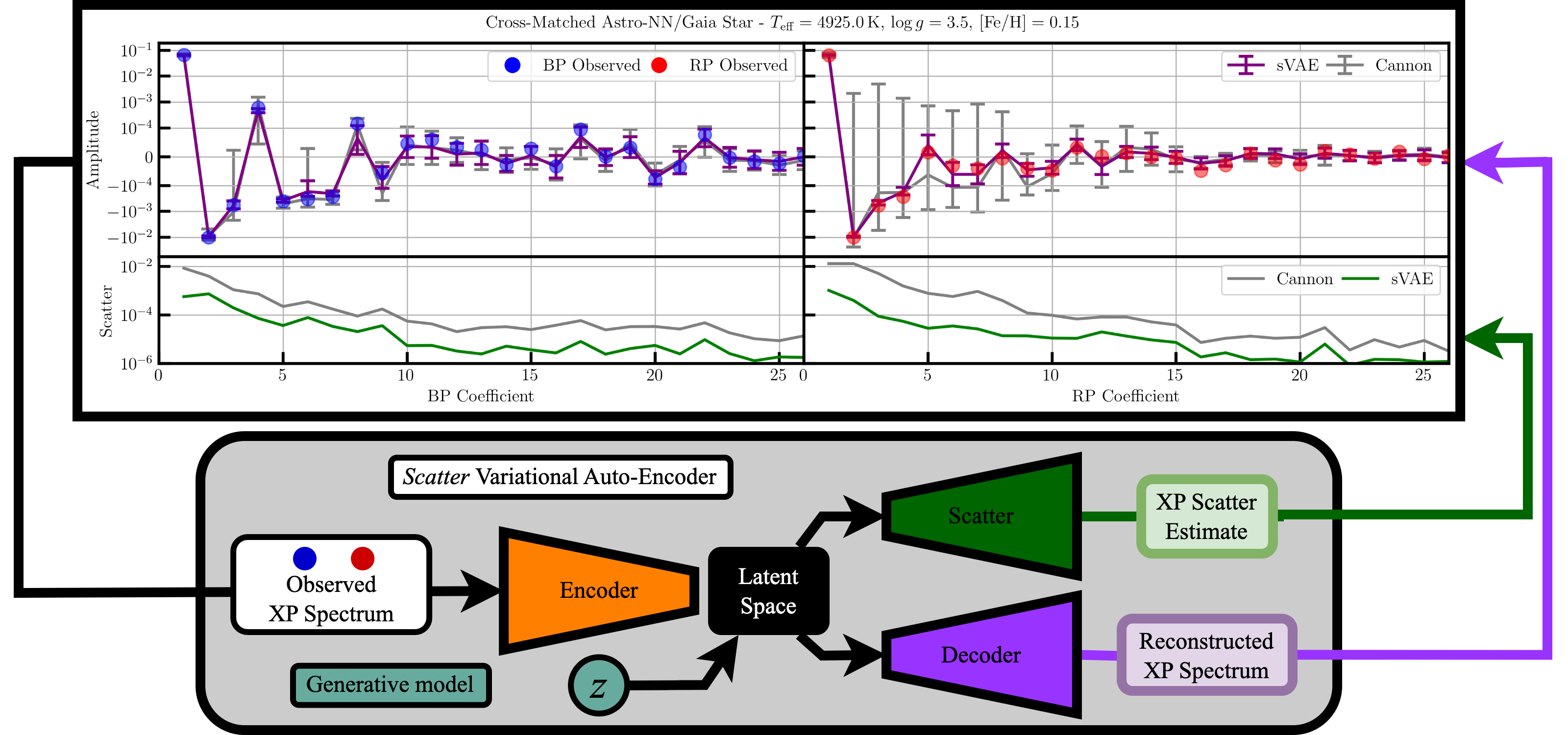}
    \caption{\emph{s}VAE archictecture with an example spectrum. An observed XP spectrum is compressed by the encoder into the latent space. The latent space then generates an XP spectrum reconstruction and a scatter estimate. We compare our \emph{s}VAE reconstruction to a \textit{Cannon} model, which overestimates scatter in comparison to the \emph{s}VAE. We do not explicitly display XP observational uncertainties as inputs since they are not propagated through the network, only factoring into our loss function (see Section \ref{ssec:train}). Note that the XP coefficient amplitudes are on a symmetric-log scale, and that error bars correspond to scatter and observational uncertainties added in quadrature.}
    \label{fig:spectra+model}
\end{figure*}

\subsection{Data} \label{ssec:xp_data}

We use the XP coefficient spectra (XPCs) in this work. Here, we outline the important characteristics of the XPCs (see \citet{GDR3-XP-22,GDR3-instrument-22} for details). \textit{Gaia} XP spectra deviate from traditional spectra in the sense that they are typically reported in Hermite polynomial space. Specifically, the BP and RP spectra are transformed from \textit{discrete} wavelength-space into \textit{continuous} coefficient-space: 110 coefficients which weight a set of Hermite polynomials (55 blue and 55 red coefficients). One can transform an XPCs into fluxes, and vice-versa, without loss of information. Indeed, \citet{Zhang++23} train their generative model in XP wavelength-space. 

Before feeding coefficient XP spectra into our model, we perform the following data pre-processing:  First, coefficients and errors are normalized by the mean flux in the \textit{Gaia} G-band, removing the brightness dependence of XP coefficients. Second, we normalize the G-band flux normalized coefficients such that each coefficient has zero mean and unit variance. An XP spectrum is accompanied by a covariance matrix. In this work, we discard off-diagonal terms of the covariance matrix, treating coefficient errors as independent. 

\subsection{\emph{Scatter} variational auto-encoder}

We present a novel implementation of a variational auto-encoder (VAE), which we term a \emph{scatter} variational auto-encoder (\emph{s}VAE). Before describing our \emph{s}VAE architecture, we briefly review the concept of a (V)AE.

An AE, which need not be variational, can be thought of as a non-linear generalization of Principal Component Analysis. An auto-encoder begins with an encoder which compresses input data, in our case a stellar spectrum, to a low-dimensional latent representation. A decoder then attempts to reconstruct the stellar spectrum from the latent representation. Ideally, this will allow the latent representation to learn key features which are shared across a set of stellar spectra. The variational nature of a VAE is added to an AE by upgrading the latent space from a collection of discrete points to a latent distribution $\mathcal{Z}$. The most popular VAE methodology is that of \citet{VAE-13}, who encode input data onto an independent multivariate Gaussian distribution. The latent space can therefore be entirely characterized by a latent mean vector $\pmb{\mu}$ and (log) variance vector $\log\pmb{\sigma}^2,$ with dimension (of the latent space) $n_{\mathcal{Z}}.$

We present the high-level architecture of our VAE in Figure \ref{fig:spectra+model}. Our \emph{s}VAE differs from a traditional VAE since, in addition to a decoder (purple) which estimates stellar spectra, we introduce a second ‘decoder’ which estimates intrinsic scatter \emph{on a star-by-star basis} (green). Importantly, both the XP reconstruction and XP scatter estimate are generated from the same latent space (black). After discarding the encoder (orange) post-training, new XP coefficients (scatter) can be generated given an arbitrary latent space vector $\pmb{z}$ (teal).

The input of our encoder is a (flux-normalized) XPCs. The XP coefficients are fed through three intermediate layers, composed of 80, 50 and 30 neurons, respectively. All intermediate layers are activated by the gaussian error linear unit (GELU). The latent parameters are then given by a linear transformation of the final intermediate layer. In this work, we have fixed the latent space to 6 dimensions, meaning the encoder produces 12 outputs (6 means and variances). Our decoder is the mirror image of our encoder, and takes as input a vector drawn from the latent distribution. We then reconstruct an XP spectrum, by feeding the latent vector through three intermediate layers analogous to the encoder, except in reverse order. Finally, a reconstruction of the 110 XP coefficients is produced with a linear transform. The scatter estimator has the exact same architecture as the decoder (while enforcing positivity). We emphasize that the weights and biases of the scatter estimator are entirely disconnected from the decoder. The scatter estimator does not produce an estimate of intrinsic scatter in the traditional sense; variance of the entire XP dataset assuming zero measurement error. Rather, the intrinsic scatter is ‘intrinsic’ to an individual star, because the scatter estimator produces different outputs on a star-by-star, or latent vector-by-latent vector basis. As such, it is more accurate to think of the intrinsic scatter as an error term which includes traditional intrinsic scatter, systematics, outliers, etc. 

\subsection{Training}\label{ssec:train}

After data pre-processing (see Section \ref{ssec:xp_data}), we train our model on ~500,000 XPCs for 100 epochs. Here, we select a subset of test Gaia XP stars which we cross-match with APOGEE to obtain stellar labels ($\sim5\times10^5$ spectra). We implement stochastic gradient descent (SGD) with momentum, with a batch size of 1024 and learning rate decay with an initial learning rate of $0.1.$ During training we aim to minimize the loss function
\begin{align}\label{eq:loss}
    \mathcal{L} = \Tilde{\chi}^2(\pmb{x},\hat{\pmb{x}},\pmb{\sigma}_x^2, \hat{\pmb{\sigma}}_s^2) &+ D_{KL}(\pmb{\mu}, \log\pmb{\sigma}^2).
\end{align}
$\Tilde{\chi}^2$ is the reconstruction loss between an input XP spectrum $\pmb{x}$ and a reconstructed spectrum $\hat{\pmb{x}}$ (from the decoder), whilst incorporating observational uncertainties $\pmb{\sigma}_x^2$ and intrinsic scatter $\hat{\pmb{\sigma}}_s^2$ (from the scatter estimator). In Eq. \eqref{eq:loss}, the $\tilde{\chi}^2$ term is given by
\begin{align}\label{eq:chi2}
    \Tilde{\chi}^2(\pmb{x},\hat{\pmb{x}},\pmb{\sigma}_x^2, \hat{\pmb{\sigma}}_s^2) =
    \chi^2(\pmb{x},\hat{\pmb{x}},\pmb{\sigma}_x^2, \hat{\pmb{\sigma}}_s^2)
    + P(\pmb{\sigma}_x^2, \hat{\pmb{\sigma}}_s^2).
\end{align}
The first term in Eq. \eqref{eq:chi2} is the traditional (reduced) $\chi^2,$ given by
\begin{align}
    \chi^2(\pmb{x},\hat{\pmb{x}},\pmb{\sigma}_x^2, \hat{\pmb{\sigma}}_s^2) =
    \frac{1}{2N}\sum_{i=1}^{N} 
    \frac{(x_i-\hat{x}_i)^2}{\sigma_{x,i}^2+\hat{\sigma}_{s,i}^2},
\end{align}
and the penalty term $P$ is given by
\begin{align}
    P(\pmb{\sigma}_x^2, \hat{\pmb{\sigma}}_s^2)
    = \frac{1}{2N}\sum_{i=1}^{N} 
    \log\left(\sigma_{x,i}^2+\hat{\sigma}_{s,i}^2\right),
\end{align}
where we are summing over the $N=110$ XP coefficients. The reconstruction loss (Eq. \eqref{eq:chi2}) can therefore be thought of as a modification to chi-square which incorporates an error term by effectively penalizing large scatter and/or uncertainties. The second term in Eq. \eqref{eq:loss} is the latent-space structure loss, for which we select the KL divergence \citep{KLD-51}, and is given by
\begin{align}\label{eq:KLD}
    D_{KL}(\pmb{\mu}, \log\pmb{\sigma}^2) = 
    \frac{1}{2n_\mathcal{Z}}\sum_{i=1}^{n_\mathcal{Z}} &
    \mu_i^2 + e^{\left(\log\sigma_i^2\right)} - \left(1+\log\sigma_i^2\right),
\end{align}
where $\pmb{\mu}$ ($\log\pmb{\sigma}^2$) are the latent means (variances) and we are summing over latent space dimensions $n_\mathcal{Z}$. In the above form, $D_{KL}$ is not particularly intuitive. For the purposes of latent space structure loss, it can be thought as a distance for probability distributions. $D_{KL}=0$ if two distributions are identical and $D_{KL}>0$ otherwise. Hence, assuming a multivariate normal latent distribution, the KL divergence is a measure of the Gaussianity of the \emph{s}VAE latent space.

\section{Results}\label{sec:res}

\begin{figure*}[hbt!]
    \centering
    \includegraphics[width=\textwidth]{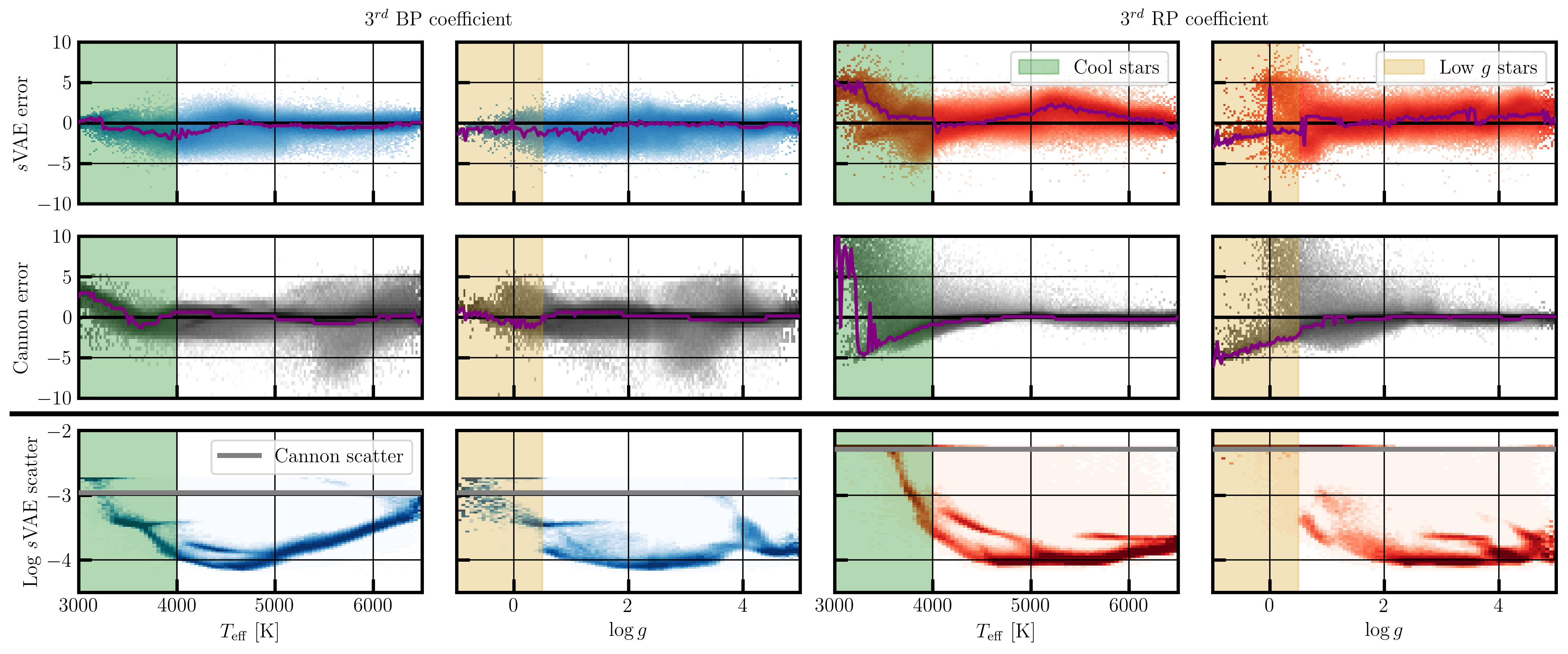}
    \caption{Normalized errors for the 3rd BP and RP coefficients, as functions of APOGEE stellar labels. \emph{s}VAE errors (\emph{top}) are compared to \emph{Cannon} model errors (\emph{middle}). \emph{s}VAE star-by-star scatter (\emph{bottom}) is compared to \emph{Cannon }population scatter (grey). We highlight the improvement in performance with our \emph{s}VAE for both cool stars (\emph{green}) and low surface gravity stars (\emph{yellow}). Purple lines indicate the mode normalized error as a function of stellar label.}
    \label{fig:errors}
\end{figure*}

We now present the results of our trained \emph{s}VAE in comparison to a \emph{Cannon} model. Reconstruction errors for our trained \emph{s}VAE (normalized by observational uncertainty and intrinsic scatter added in quadrature) are presented in Figure \ref{fig:errors}. The \emph{Cannon} model reconstructs an XP spectrum from stellar labels. We train the \emph{Cannon} model on the same APOGEE/XP cross-match used for \emph{s}VAE training in Section \ref{ssec:train}. Our unsupervised \emph{s}VAE has three key advantages over existing supervised generative models: 
\begin{enumerate}
    \item Star-by-star intrinsic scatter estimation (and overall model flexibility) yields better XP reconstructions over all stellar parameter space.
    \item Our trained \emph{s}VAE accurately reconstructs stellar spectra in regions of stellar parameter space where supervised learning failed with (bad) stellar labels.
    \item Our model covers far more of the XP parameter space than supervised learning.
\end{enumerate}
First, in Figure \ref{fig:errors}, normalized errors for the \emph{s}VAE clearly truncate at $\sim5\sigma,$ whereas the \emph{Cannon} errors can extend beyond $10\sigma.$ Although we only present the 3rd XP coefficients in Figure \ref{fig:errors}, the \emph{average}, normalized errors for our trained \emph{s}VAE across all XPCs in the APOGEE cross-match are only $2\%$ ($9\%$) larger than the \emph{Cannon} model, for the first 10 BP (RP) coefficients. This is more impressive if we compare individual XP scatter to \emph{Cannon} population scatter in the bottom row of Figure \ref{fig:errors}. The \emph{s}VAE achieves the same accuracy as the \emph{Cannon} model, with respect to normalized errors, while simultaneously having far smaller intrinsic scatter estimates (by up to 2 order of magnitude). The \emph{Cannon} model overestimates intrinsic scatter, and by extension reduces normalized errors. In contrast, the lower (individual) scatter estimates of the \emph{s}VAE allow an overall increase in reconstruction accuracy. In summary, our generically better XP reconstructions, relative to the \emph{Cannon} model, can be attributed to $(i)$ inclusion of individual scatter estimation and $(ii)$ the flexibility of the generative function which relates the latent space to XPCs.

Second, the lack of reliance on stellar labels in our unsupervised approach is a significant advantage over supervised learning at the extrema of stellar parameter space. In Figure \ref{fig:errors}, we observe that our model is less susceptible to large errors at both the cool end of $T_{\rm eff}$ (\emph{green}) and the low $g$ tail (\emph{yellow}). In contrast, \emph{Cannon} model errors blow up because supervised learning fails when bad stellar label estimates are used. Our \emph{s}VAE avoids this shortcoming by swapping out stellar labels for a latent space as its generator.

\begin{figure}[hbt!]
    \centering
    \includegraphics[width=\columnwidth]{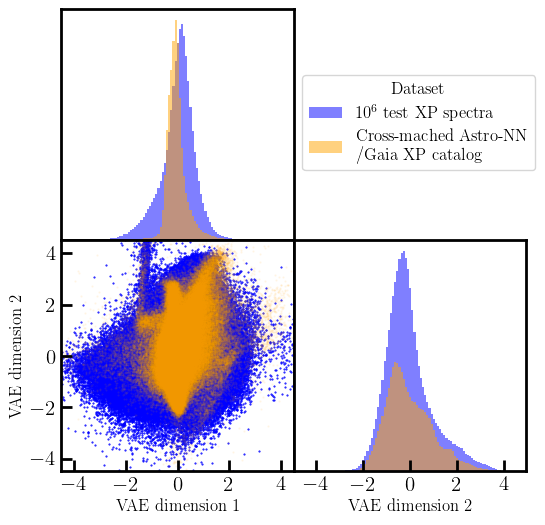}
    \caption{Latent space projection for the APOGEE/XP cross-matched catalog (orange) and $10^6$ unlabeled XP spectra (blue).}
    \label{fig:latent}
\end{figure}

Third, our unsupervised approach greatly increases ‘coverage’ of the XP parameter space, relative to supervised learning. The XP spectra contain rare stellar populations which are not well characterized by the current stellar label catalogs available for supervised learning. To illustrate this, we project both the entire XP/APOGEE cross-matched catalog and $10^6$ XP spectra (outside of the cross-match) into the \emph{s}VAE latent space. This is visualized across two latent space dimensions in Figure \ref{fig:latent}. We observe that the XP/APOGEE cross-match covers significantly less latent space volume than the $10^6$ XP spectra without APOGEE labels. We compute the relative latent space volume difference by numerical integration and find $\Delta V/V_{\rm sup}\approx 30\%,$ meaning the $10^6$ unlabeled spectra cover 30\% more latent space than the APOGEE cross-match. Extrapolating this to the entire 220 million XP spectra, our \emph{s}VAE improves latent space coverage by approximately a factor of $220\times\frac{\Delta V}{100V_{\rm sup}}\approx65$. We emphasize that this increase in latent space coverage is a crude approximation of the ‘breadth’ gained by our unsupervised approach. First, latent space coverage does not necessarily correspond to better stellar parameter space coverage (although it certainly can). Second, the \emph{s}VAE is degenerate in the sense that the same model can learn different embeddings of the data, due to latent space symmetries. Therefore, the 65x factor we quote above is only qualitative. It serves to motivate the use of our \emph{s}VAE: better latent space coverage, by roughly an order of magnitude, is a proxy for better information coverage over the XP dataset.

\section{Conclusion \& Future Work}\label{sec:disc}

We developed a novel deep, generative model: a \emph{scatter} variational auto-encoder, and applied it to the \emph{Gaia} XP spectra. We showed that our \emph{s}VAE outperforms supervised learning (\emph{Cannon}) in many respects. Most importantly, our unsupervised approach has the potential to close the stellar labels gap. Firstly, for XP spectra with stellar labels from previous spectroscopic surveys, our \emph{s}VAE produces better reconstructions than supervised learning if stellar labels are poorly estimated. Second, our unsupervised model can cover the entire XP parameter space, whereas supervised learning is limited to only a small subset, because we swap out stellar labels for latent variables. 

The evident shortcoming of trading stellar labels for latent labels is the inability to perform stellar label inference with our unsupervised model. Nevertheless, inference can be performed in the latent space to identify \emph{relative} differences between stellar populations. Our \emph{s}VAE, combined with existing supervised models \citep[e.g.][]{Zhang++23}, will eventually yield an accurate, \emph{semi}-supervised generative model for the entire \emph{Gaia} XP data. Our approach is also promising for both outlier detection and identification of rare stellar populations. For example, we expect our approach will provide additional insights into the CEMP star catalog of \citet{Lucey++22} and aim to pursue this in future work.

\bibliography{refs}
\bibliographystyle{icml2023}


\end{document}